\documentclass[a4paper,11pt]{article}
\usepackage{pos}
\begin{document}
\title{The deployment of the ICARUS LAr detector on the  short baseline 
(SBN) neutrino beam at FNAL}

\ShortTitle{The Icarus T600 detector at FNAL}

\author[a,1]{M. Bonesini}
\author[b,1]{A. Menegolli}
\affiliation[a]{Sezione INFN Milano Bicocca, \\
        Dipartimento di Fisica G. Occhialini, 
        Universit\'a di Milano Bicocca, Milano, Italy}
\affiliation[b]{Sezione INFN Pavia, \\
        Dipartimento di Fisica, Universit\'a degli Studi di Pavia, 
	Pavia, Italy}  
\emailAdd{Maurizio.Bonesini@mib.infn.it}
\emailAdd{Alessandro.Menegolli@pv.infn.it}
        \note{ {\bf on behalf of the Icarus Collaboration}} 

\abstract{
The Icarus T600 detector represents the first example of a 
fully working large-mass LAr detector.  After operations at the LNGS INFN
laboratories, it has been refurbished at CERN in 2015-2017 and
then installed as far detector on the BNB neutrino beamline at FNAL. The main
operations involved in the T600 overhauling are thouroghly described  in this paper.} 

\FullConference{
European Physical Society Conference on High Energy Physics - EPS-HEP2019 -\\
			10-17 July, 2019\\
			Ghent, Belgium}

\maketitle
\section*{Introduction}

The ICARUS T600 detector is made of two identical modules for a total mass of $\sim$ 760 tons of Liquid Argon, representing the biggest detector of this kind in operation. Each module is equipped with two readout chambers on the long sides, with planes of wires at 0$^\circ$, $\pm$60$^\circ$ for a total of 54000 readout wires.
ICARUS T600  has been previously installed in the underground INFN-LNGS 
Gran Sasso Laboratory and has been the first large-mass LAr TPC operating 
as a continuously sensitive general purpose observatory \cite{icarus}. 
The operation of the ICARUS T600 LAr TPC demonstrated the enormous potential 
of this detection technique, addressing a wide physics program with the 
simultaneous exposure to the CNGS neutrino beam and cosmic rays.
The ICARUS T600 detector is currently being deployed at FNAL as a far detector of the Short-Baseline Neutrino (SBN) program \cite{sbn}, dedicated to clarify the sterile neutrino anomalies by precisely and independently measuring both $\nu_e$ appearance and $\nu_\mu$ disappearance in the FNAL Booster Neutrino Beam (BNB).

\section*{The CERN overhauling phase}

\begin{figure}
\begin{center}
\includegraphics[width=.32\textwidth]{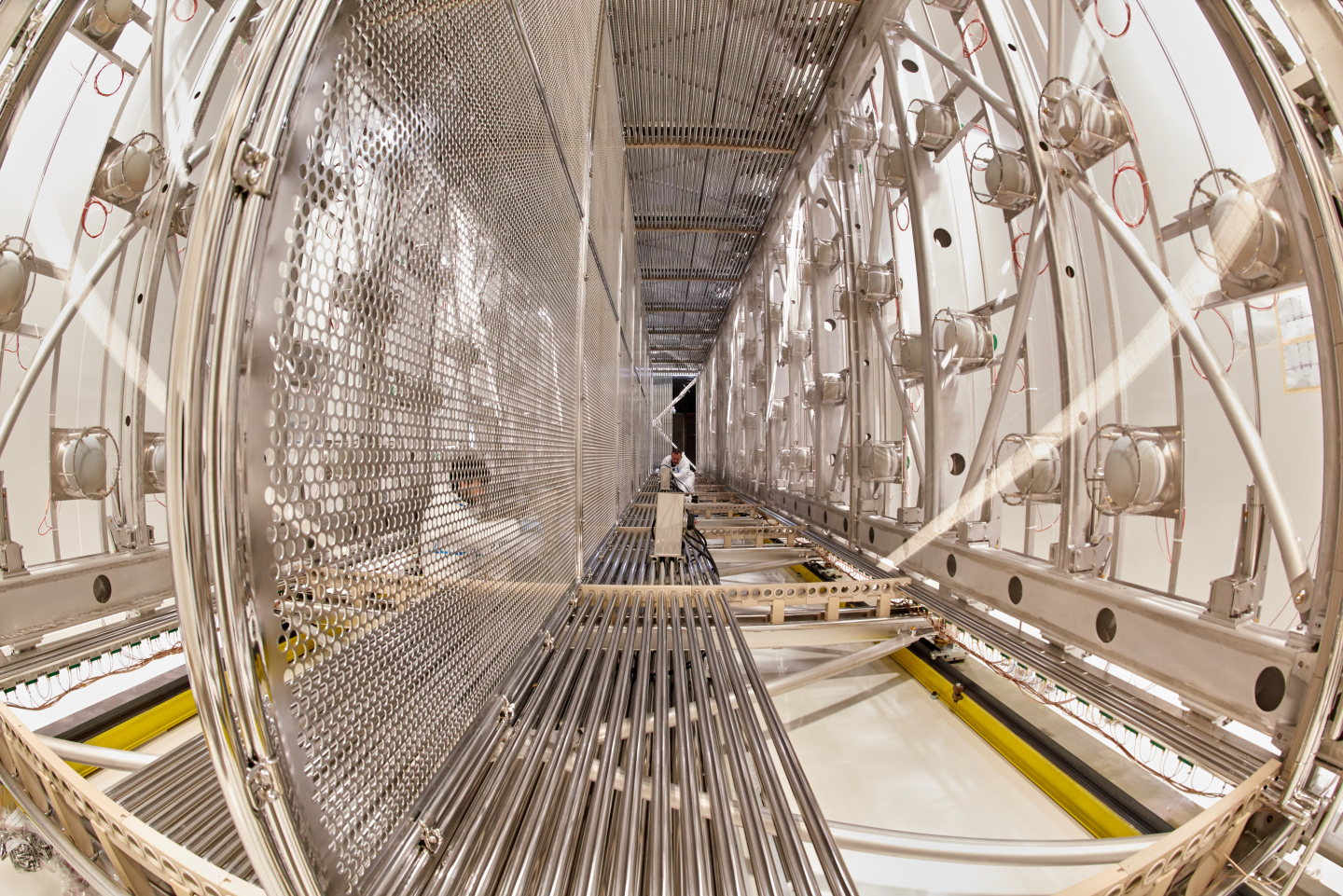}
\includegraphics[width=.26\textwidth]{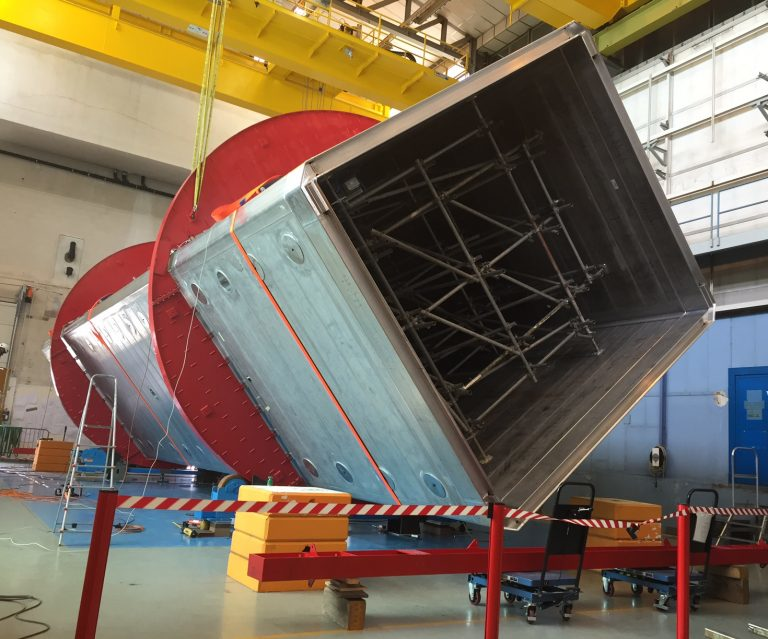}
\includegraphics[width=.36\textwidth]{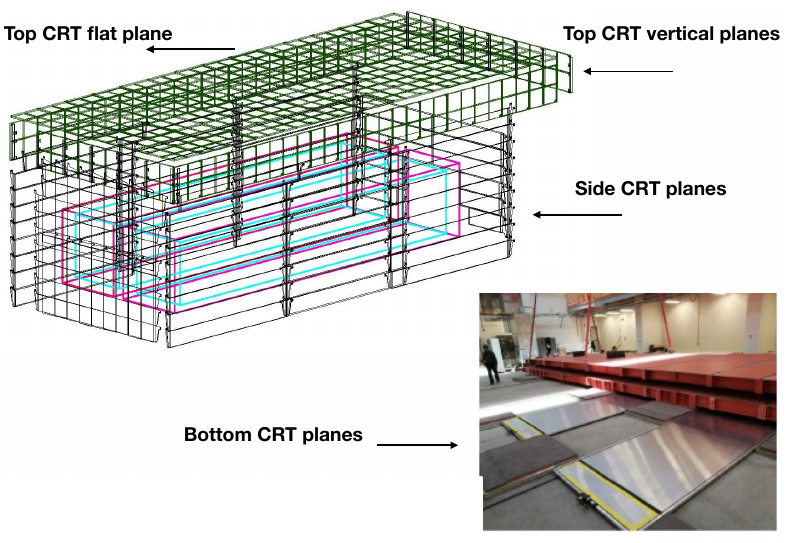}
\end{center}
\caption{Left: inner view of one ICARUS module with the two TPCs separated by 
the common central cathode and the new 8" PMT system behind the wire planes,
during refurbishing in a CERN clean room. 
Center: one of the new cold vessels in a CERN workshop. 
Right: sketch of the layout of the Icarus CRT and picture of 
bottom 
CRT panels.}
\label{wa104}
\end{figure}

Having to face a more severe experimental condition than at the LNGS 
underground site, due 
to the presence of a 11 kHz cosmic rays background, the ICARUS T600 detector 
underwent an intensive overhauling at CERN in the Neutrino Platform 
framework from 2015 to 2017, before being shipped to FNAL. 
Several improvements were introduced, while maintaining 
the already achieved good performance obtained during the  LNGS run:

\begin{itemize}
\item new cold vessels, with a purely passive insulation;
\item renovated LAr cryogenics/purification equipment;
\item improvement of the cathode planarity;
\item upgrade of the light scintillation detection (PMT) system with higher granularity and a time resolution
around 1 ns;
\item new faster, higher-performance read-out electronics for the wire chambers.
\end{itemize}
To handle  the large flux of cosmic rays,
     a 3 m concrete overburden to remove contribution from charged 
hadrons/$\gamma$'s and
a 4$\pi$ external Cosmic Ray Tagger (CRT) to correlate residual muons 
with TPC signals were introduced 
in addition to the PMT system.
Some pictures of the above improvements and the overhauling phase at CERN are
shown in figure \ref{wa104}.

\section*{The ICARUS deployment at Fermilab}

ICARUS modules were placed inside the warm vessel in August 2018 after their shipping to FNAL in 2017.
Installation of the TPC and PMT feed-through flanges and connectivity tests were completed by February 2019.
In sequence leak tightness tests were then performed and
top cold shields and top CRT support have been installed.
Side CRT installation is also ongoing.
The installation of proximity cryogenics started in February 2019 
and is almost completed.
Evacuation of the detector started on June 2019 to reach  a residual internal pressure of $\sim$ 10$^{-4}$ mbar.

\begin{figure}
\begin{center}
\includegraphics[width=.8\textwidth]{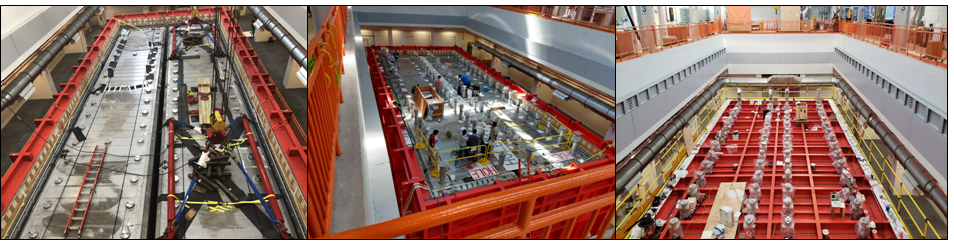}
\end{center}
\caption{Left: placement of the two ICARUS modules 
inside the warm vessel  in the far detector building at Fermilab - August 2018. Center: chimneys installation - October 2018; Right: feed-through installation -December 2018.}
\label{installation1}
\end{figure}

\begin{figure}
\begin{center}
\includegraphics[width=.6\textwidth]{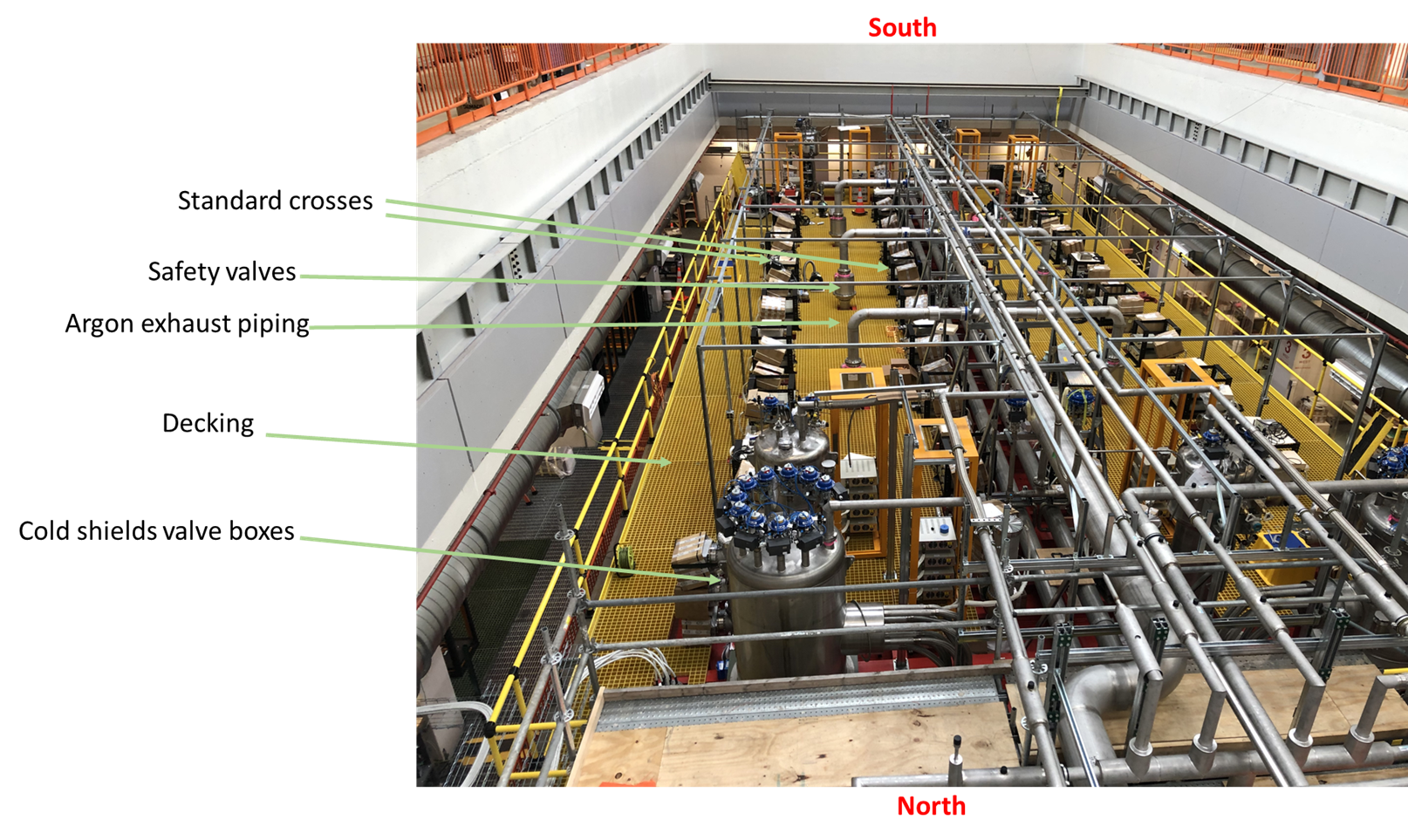}
\end{center}
\caption{Present status of the ICARUS cryogenic plant.}
\label{installation2}
\end{figure}

Some pictures related to these deployment operations at FNAL are shown 
in Figure \ref{installation1}, while the present status of the cryogenic plant 
is shown in figure \ref{installation2}.

\section*{The new Icarus light detection system.}

\begin{figure}
\begin{center}
\includegraphics[width=.36\textwidth]{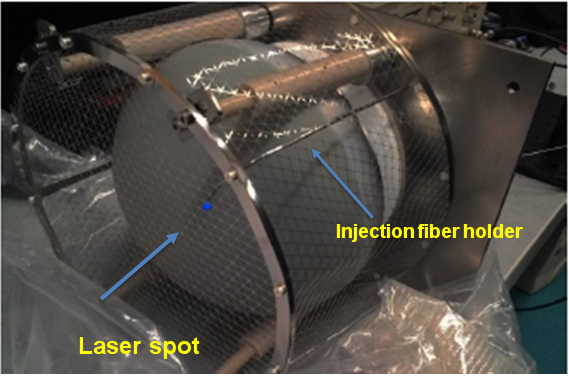}
\includegraphics[width=.36\textwidth]{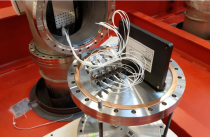}
\end{center}
\caption{Left: picture of the laser injection system onto the PMT window. 
Right: installation of one flange housing the optical feedthrough for 
laser calibration. The $1 \times 10$ optical splitter stands up on the right, 
while a patch panel to connect the splitter pigtails to the 7 m long injection
fibers is shown on the left.}
\label{laser1}
\end{figure}

Scintillation in LAr is characterized by a prompt photon emission in
the VUV region at $\lambda \sim 128$ nm with a yield $\sim 1-2 \times 10^{-4}$
$\gamma$'s/MeV. Detection is done through large area  8`` PMTs with a TPB
coated glass window, acting as a wavelength shifter to the PMT photocathode 
sensitive region.
The new light collection system will allow to precisely identify the time 
of occurrence ($t_0$) of any ionizing event in the ICARUS TPCs, determine the 
event rough topology for selection purposes and generate a trigger signal for read-out.
The PMT system  exploits  90 8" Hamamatsu R5912-MOD PMTs per TPC (5\% coverage, 15 phe/MeV) providing sensitivity to low energy events ($\sim$ 100 MeV), good spatial resolution ($\leq$ 50 cm) and  $\sim 1$ ns timing resolution \cite{pmt}.
The readout is via CAEN V1730B digitizers (500 Ms/s, 14 bit, 2 Vpp dynamic
range). The time evolution of the PMT timing/gain equalization will be traced 
by using fast light pulses from a Hamamatsu PLP10 diode laser source 
(FWHM $\sim$ 60 ps, 405 nm, 120 mW power).
Laser pulse are sent to each PMT via a distribution system based on an 
Agiltron optical switch (1x46, only 36 channels used) connected to  36 
20 m long multimode (MM) 50 $\mu$m   core fiber patch cords from OZ/Optics, 
going to 36 CF40 to CF200 adapters holding each one an optical feedthrough
from Vacom Gmbh. Each adapter is equipped with one $1 \times 10$ fused fibers 
optical splitter from
Lightel US on the internal side. 
 Inside the cryostat 7 m long optical injection fibers connected
to the output pigtails of the $1 \times 10$ splitters convey the light in front of each PMT.
A picture of the injection fiber holder in front of one PMT and of one
 CF40 to CF200  
adapter  is shown in figure \ref{laser1}.

\begin{figure}
\begin{center}
\includegraphics[width=0.7\textwidth]{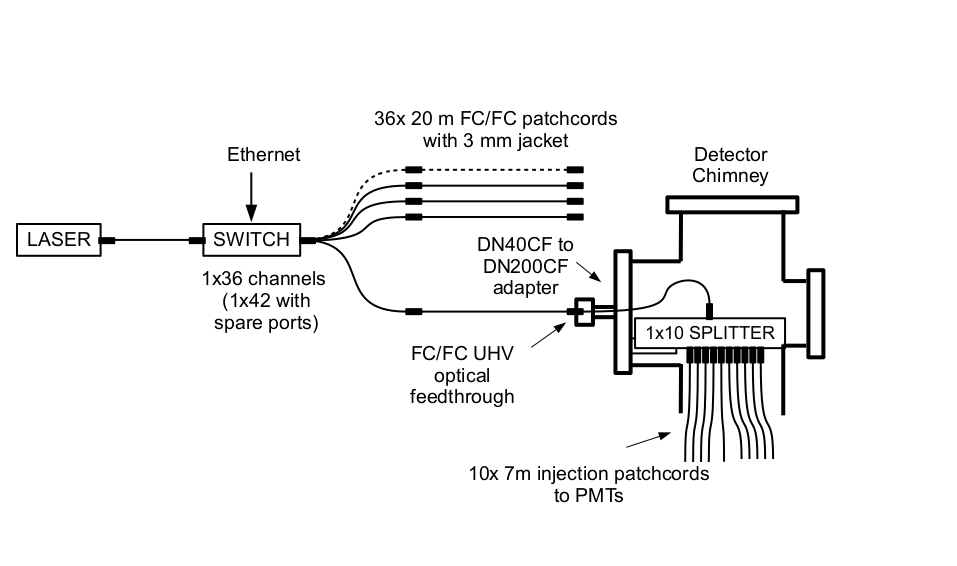}
\end{center}
\vskip -1cm
\caption{Schematic layout of the laser calibration system.}
\label{laser}
\end{figure}

The layout of the calibration system is shown in figure \ref{laser}.
In addition the system will include also a fast photodiode (Thorlabs
DET02AFC) for monitoring the laser stability and a remotely controlled
system to attenuate the input laser power to the delivery system. 

All PMTs and  calibration system components parts were individually tested.
In particular:
\begin{itemize}
\item All PMTs were tested at room temperature in a dedicated dark room at CERN. A subset of 60 PMTs tested was immersed in LAr to compare the PMT performance in cryogenic environment to room temperature;
\item all optical patch cords , 1x10 splitters, UHV optical 
flanges and internal injection fibers  have been characterized in terms 
of time spread, transmission and rise time with a 20 GHz sampling 
oscilloscope \cite{laser}, after preliminary tests to select the most 
suitable components \cite{laser1};
\item the laser diode  and the optical switch stabilities were tested 
over a period of many months.
\end{itemize}
All 108 feed-through flanges (36 for PMT HV, 36 for PMT signal and 36 for optical fibers) were installed from December 2018 to February 2019.

\section*{First results on the ICARUS wire read-out}

\begin{figure}
\begin{center}
\includegraphics[width=.32\textwidth]{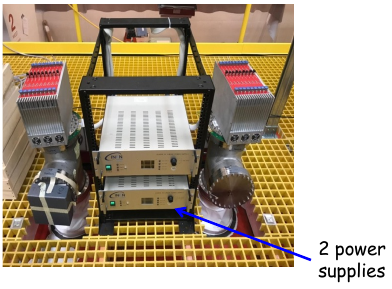}
\includegraphics[width=.46\textwidth]{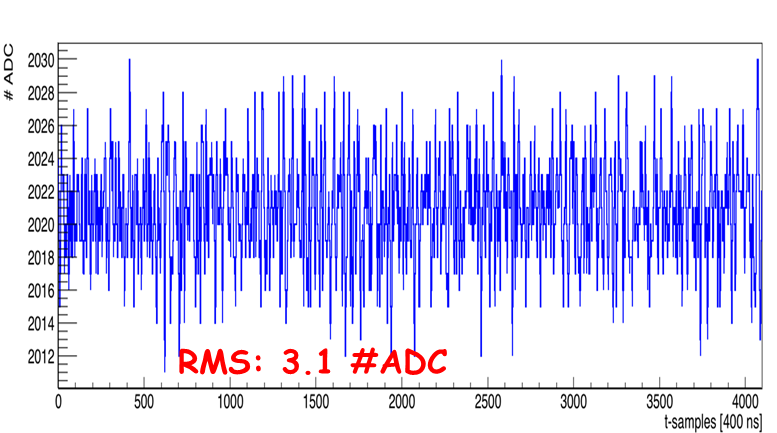}
\end{center}
\caption{Left: two mini-crates with 9 A2795
boards on two chimneys for Induction 2 and Collection wires
connected to power supplies. Right: example of wire response to test pulses for noise characterization.}
\label{tpc}
\end{figure}

The new ICARUS TPC electronics \cite{electronics} has a  front-end based 
on analogue low noise/charge sensitive pre-amplifiers, a serial 12 bit ADC, one per channel with 400 ns sync. sampling and a serial bus architecture with optical links for Gigabit/s transmission.
Analogue/digital electronics is directly mounted on the signal 
feedthrough flanges: the external side of
flange itself is the back-plane for a custom mini-crate, see figure \ref{tpc} 
for some details.
The read-out is performed by nine CAEN A2795 boards per chimney 
housed in a mini-crate serving 576 channels.
Each board hosts 64 pre-amplifiers, 64 12 bit, 2.5 Ms/s  ADCs, 
 FPGA, memory and  optical link interface. From the output of the preamplifier,
the board operates as a waveform digitizer. The back-plane of 
a mini-crate distributes power supply and local control signals.
All feed-through flanges and mini-crates with the TPC wire read-out electronics (576 channels + optical links) have been installed. A successful test of the full readout chain, from wires to DAQ, was performed in April/May 2019 for all the mini-crates.

An example of the noise determination for wires is shown in the right panel 
of figure \ref{tpc}.
\section*{Conclusions}
The Icarus T600 detector is being installed at Fermilab, as Far Detector 
on the BNB beamline. It has been successfully upgraded at Cern, within the
WA104 project framework and will be soon ready for data taking at Fermilab.

\end{document}